\documentclass[11pt,twoside]{article}
\usepackage{./asp2014}

\aspSuppressVolSlug
\resetcounters

\begin{document}

\title{Pulsational Mass Ejection in Be Star Disks}
\author{Nathaniel Dylan Kee,$^1$ Stan Owocki,$^1$ Richard Townsend,$^2$ and Hans-Reinhard M{\"u}ller$^3$}
\affil{$^1$University of Delaware, Newark, DE, USA; \email{dkee@udel.edu}}
\affil{$^2$University of Wisconsin-Madison, Madison, WI, USA}
\affil{$^3$Dartmouth College, Hanover, NH, USA}

\paperauthor{Nathaniel Dylan Kee}{dkee@udel.edu}{ORCID_Or_Blank}{University of Delaware}{Physics and Astronomy}{Newark}{DE}{19717}{USA}
\paperauthor{Stanley Owocki}{owocki@bartol.udel.edu}{ORCID_Or_Blank}{University of Delaware}{Physics and Astronomy}{Newark}{DE}{19717}{USA}
\paperauthor{Richard Townsend}{townsend@astro.wisc.edu}{ORCID_Or_Blank}{University of Wisconsin-Madison}{Astronomy}{Madison}{WI}{53706}{USA}
\paperauthor{Hans-Reinhard M{\"u}ller}{}{ORCID_Or_Blank}{Dartmouth College}{Physics and Astronomy}{Hanover}{NH}{03755}{USA}

\begin{abstract}
This work explores a Pulsationally Driven Orbital Mass Ejection (PDOME) model for the launching of Classical Be star disks. Under this model, a combination of rapid rotation and non-radial pulsation modes contribute to placing material into the circumstellar environment. Several varieties of non-radial pulsation modes, characterized by their propagation direction and the relative phase of their velocity and density perturbations, are considered. As well, the orbital stability of material launched by such a mechanism is investigated.
\end{abstract}

\section{Introduction}

Despite the long and extensive history of observations of Classical Be stars, much remains unknown about the physics that forms their circumstellar disks. Of key interest is the mechanism by which such stars are able to launch material into their circumstellar environment with enough angular momentum to form a disk. At this point, it is widely accepted that the universal appearance of rapid rotation in this class of objects plays a central role in this process. Though exact rotation velocities for individual Be stars are not well agreed upon due to the flattening of the relationship between observed line broadening and stellar rotation velocity in the limit of rapid rotation \citep{TowOwo04}, for sub-critical rotation, some additional mechanism(s) are necessary to place material into orbit.

A crucial step toward an understanding of the origins of the circumstellar material came from the wind compressed disk (WCD) model of \cite{BjoCas93}. In this model, the radial velocity contribution of line-driving in the stellar wind, coupled with the tangential velocity component from the rapid rotation of the star, places material in the wind into inclined orbits that eventually pass through the equatorial plane, whereupon collision with material from the opposite hemisphere forms an equatorial WCD. However, the work of \cite{OwoCra96} showed that the combination of non-radial forces and the equatorial gravity darkening present in rapidly rotating, oblate stars combine to reverse the expected equatorward flow into a poleward flow, thereby inhibiting the formation of an equatorial disk.

A possible clue to an alternate mechanism came from observations of line profile variability of $\mu$ Cen by \cite{RivBaa01} and $\omega$ CMa by \cite{MaiRiv03}. In both cases, increases in the amplitude of photometric variability from non-radial pulsations were found to be associated with modulation and growth of emission from the circumstellar disk. Since then, observations have shown that non-radial pulsation in Classical Be stars is a ubiquitous feature, and \cite{RivBaa03} have even gone so far as to suggest that these non-radial pulsations may be causally connected with the Be phenomena.

The Pulsationally Driven Orbital Mass Ejection (PDOME) model here explores the dynamical issues for ejecting material into orbit from non-radial pulsations on a star near critical rotation. Section \ref{model} discusses details of the model and the parameters used, section \ref{results} presents some preliminary findings, and section \ref{conclusions} discusses possible fruitful directions for future work.

\section{Details of the Model}\label{model}

\subsection{Basic Paradigm}

Our approach here is to explore the dynamics of circumstellar material launched within the equatorial plane from perturbations in density and azimuthal velocity on an underlying, rapidly rotating Be star. Before beginning a discussion of the model in full, it is helpful first to define and clarify some terms. The most important among these are concerned with the nature of the surface perturbations intended to mimic non-radial gravity waves ($g$-mode) pulsations.

For computational convenience, the perturbations explored here are always assumed to be $\lvert m\rvert=4$ , where $m$ is the number of nodes around the equator of the star; however, our implementation of lower boundary conditions in density and velocity allows for distinct directions of phase propagation and energy transport. Specifically, in the frame of the star, the \emph{phase} propagation can either be \emph{prograde} ($v_{phase}>0$) or \emph{retrograde} ($v_{phase}<0$) relative to the direction of stellar rotation. Separately, we can also force the \emph{material} velocity at the peak density of the perturbation to be either prograde ($v_{pert}(\rho_{max})>0$) or retrograde ($v_{pert}(\rho_{max})<0$). Since the material motion sets the direction of energy propagation that is normally associated with group velocity, we refer to the four combinations we explore as prograde group/prograde phase (+/+), prograde group/retrograde phase (+/-), etc.

\subsection{Numerical Model Specifications}

Our numerical simulations of the PDOME model use the Piecewise-Parabolic Method \citep{ColWoo84} hydrodynamics code\footnote{ http://wonka.physics.ncsu.edu/pub/VH-1/} VH-1, implemented here in a 2D, spherical equatorial ($r,\phi$) plane with azimuth ranging from 0 to 90 degrees and radius from 1 to 6 $R_*$. For simplicity we use an isothermal approximation without explicit inclusion of viscous terms.

As the scale height of the stellar atmosphere is very small, $H\sim\,R_*/1000$, it is difficult to resolve stellar pulsations within a hydrodynamic simulation focused on the dynamics of a circumstellar disk over several stellar radii. We thus, instead, mimic the effect of pulsations by imposing sinusoidal perturbations in density and azimuthal velocity at the lower boundary,
\begin{align}
\rho(\phi) &= \rho_0 10^{\left(\log\left(\frac{\rho_{max}}{\rho_0}\right)\sin\left(\frac{2 \pi t}{P}+m\phi\right)\right)} \label{d_pert}\\
v_\phi(\phi) &= v_{rot} + v_{\phi,pert}\sin\left(\frac{2\pi t}{P}+m\phi+\phi_0\right), \label {w_pert}
\end{align}
where the exponential variation in density reflects the exponential stratification of the stellar atmosphere, with mean density $\rho_0$ and maximum density of the perturbation $\rho_{max}$. Here $v_{\phi,pert}$ is the azimuthal velocity perturbation, $P$ is the perturbation period, and $\lvert m \rvert$ is the number of nodes around the stellar equator, with $m>0$ ($m<0$) giving prograde (retrograde) phase velocity. For $\phi_0 = 0^\circ$, the perturbations in density and and velocity are in phase, representing a prograde group velocity; for $\phi_0=180^\circ$ the velocity and density perturbations are in antiphase, signifying retrograde group velocity. Table \ref{vp_and_vg} summarizes the 4 intercombinations explored in the models detailed in section \ref{results} and what each implies for phase and group velocity.

\begin{table}
\begin{center}
\def\arraystretch{1.75}
\begin{tabular}{r|c|c}
~&$m>0$&$m<0$\\
\tableline
$\phi_0=0^\circ$&$v_p>0$,$v_g>0$&$v_p<0$,$v_g>0$\\
\tableline
$\phi_0=180^\circ$&$v_p>0$,$v_g<0$&$v_p<0$,$v_g<0$
\end{tabular}
\end{center}
\caption{Sense of phase and group velocity as a function of the sign of $m$ and the value of $\phi_0$.}
\label{vp_and_vg}
\end{table}

Apart from the variations noted in table \ref{vp_and_vg}, all models share common parameters (table \ref{params}). Parameters are chosen to be roughly representative of those inferred for a typical pulsating Be star (eg. $\mu$ Cen). For simplicity, $M_*$ and $R_*$ are tuned to give an equatorial surface orbital speed $v_{orb}=500\;\mathrm{km}\,\mathrm{s}^{-1}.$ This fixed value for all models allows for the introduction of the stellar parameter $W\equiv v_{rot}/v_{orb}$, where $v_{rot}$ is the equatorial rotation speed. For the standard set of parameters, $W=0.95$ and $v_{rot}=v_{orb}-c_s$ where $c_s= 25\;\mathrm{km}\,\mathrm{s}^{-1}$ is the sound speed.

\begin{table}[!ht]
\begin{center}
\def\arraystretch{1.75}
\begin{tabular}{|l|c|}
\multicolumn{2}{c}{Stellar Parameters}\\
\tableline
$M_*$&9.2$M_\odot$\\
\tableline
$R_*$&7$R_\odot$\\
\tableline
$v_{rot}$&$475\;km\,s^{-1}$\\
\tableline
$c_s$&$25\;km\,s^{-1}$\\
\tableline
\end{tabular}
\hspace{10pt}
\begin{tabular}{|l|c|}
\multicolumn{2}{c}{Perturbation Parameters}\\
\tableline
$P$&40\;ks\\
\tableline
$|m|$&4\\
\tableline
\end{tabular}
\end{center}
\caption{Standard simulation parameters.}\label{params}
\end{table}

\section{Results}\label{results}

\subsection{Prograde vs. Retrograde Phase and Group Velocity}

\articlefigure{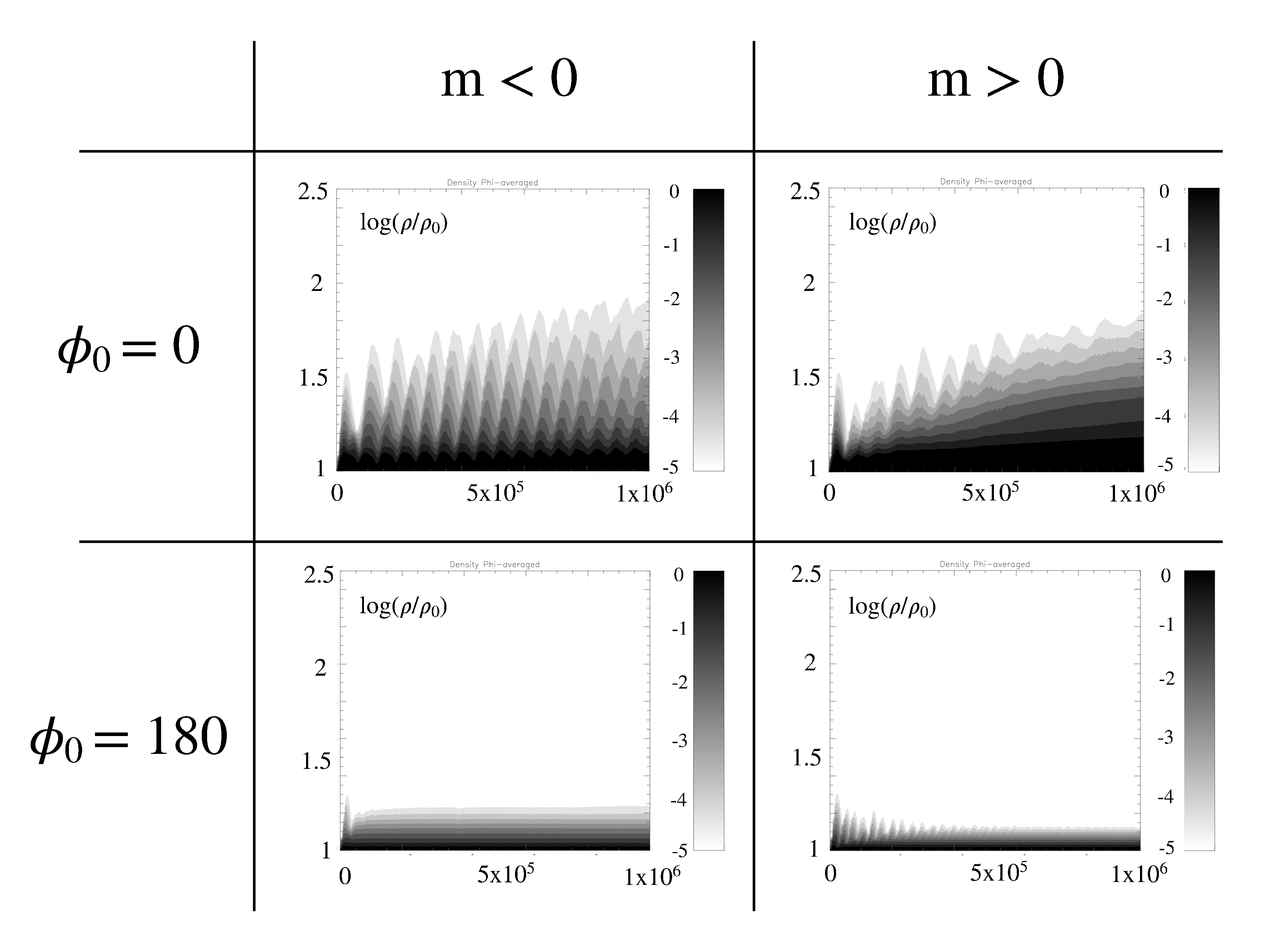}{logrho_comp}{$\log(\rho)$ in $\mathrm{g}\,\mathrm{cm}^{-3}$ as a function of radius, measured in $R_*$ and time in seconds. While all combinations put some mass into orbit, prograde group velocity simulations do so much more efficiently than those with retrograde group velocity.}

In considering results, let us first compare the behavior of the four possible combinations of phase and group velocity listed in table \ref{vp_and_vg}. Figure \ref{logrho_comp} compares results for the log of the azimuthally averaged density, $\log\langle\rho\rangle_\phi$, as a function of radius and time. While the outer boundary of the simulation is at $6R_*$, the figures focus on the most interesting behavior, below $2.5R_*$. Figure \ref{mid} plots the total mass in the disk as a function of time, computed by summing density over the radial direction, while accounting for mass that escapes through the outer boundary. Note that, while all four models do put some material into orbit, prograde group velocity models do so much more effectively. Moreover, observations favor a retrograde phase velocity model \citep{RivBaa03}, thus our favored model is a mixed phase model with prograde group velocity and retrograde phase velocity, henceforth referred to as the +/- model.

\articlefigure{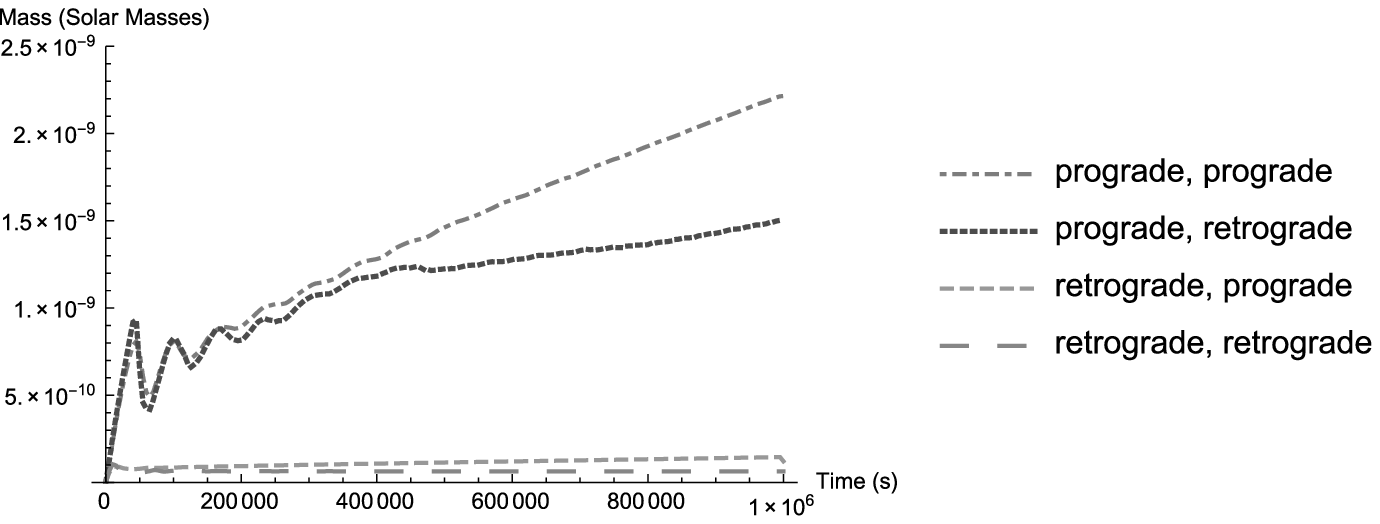}{mid}{Mass in the disk in $M_\odot$ as a function of time. Sense of velocity is listed with group first, then phase.}

\subsection{+/- model}

\articlefigure{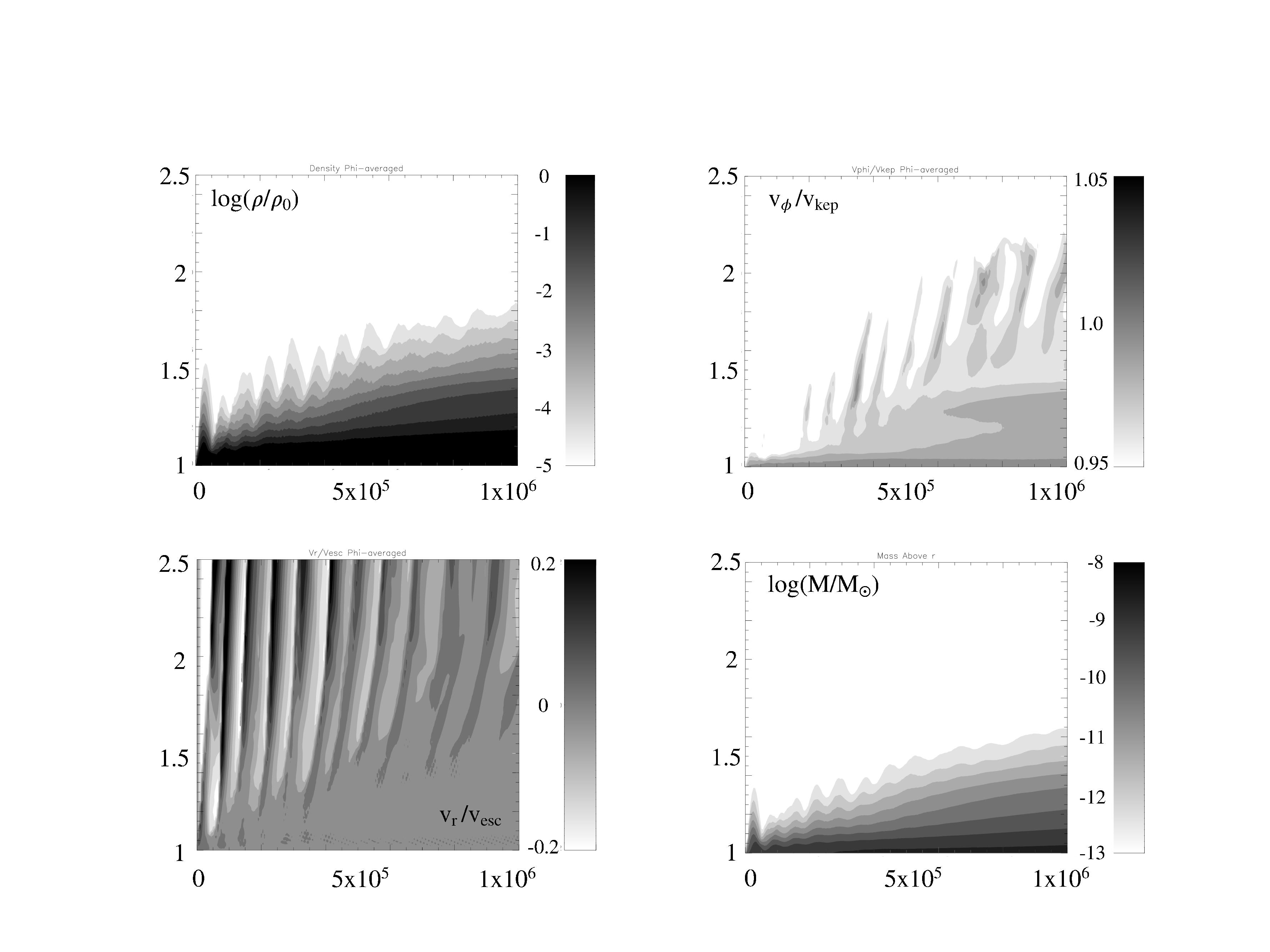}{pra}{Clockwise from the upper left, $\log(\rho)$, Kepler number, mass above each radius in $M_\odot$, and radial velocity in units of escape velocity for the +/- model as a function of radius and time.}

For considering a PDOME model in detail, there are four quantities of particular merit, namely: $\log\langle\rho\rangle_\phi$; mass above each radius; radial velocity; and Kepler number, the ratio of azimuthal velocity to local Keplerian orbital velocity. These are plotted in figure \ref{pra} for the +/- model. The first of these, $\log\langle\rho\rangle_\phi$, has already been shown and discussed for all models in figure \ref{logrho_comp}.  The mass above each radius shows that the mass launched into orbit is comparable to inferred total masses in Classical Be disks.
Radial pressure support allows the disk to be nearly in hydrostatic equilibrium in the radial direction for the inner disk even with Kepler number slightly below unity and, indeed, the lower left panel of figure \ref{pra} shows that radial velocity is significantly below the local escape speed.
Figure \ref{rel} shows that, if the pulsation is turned off, about half of the material begins to settle into a stable disk, while the remainder falls back on to the star. 

\articlefigure{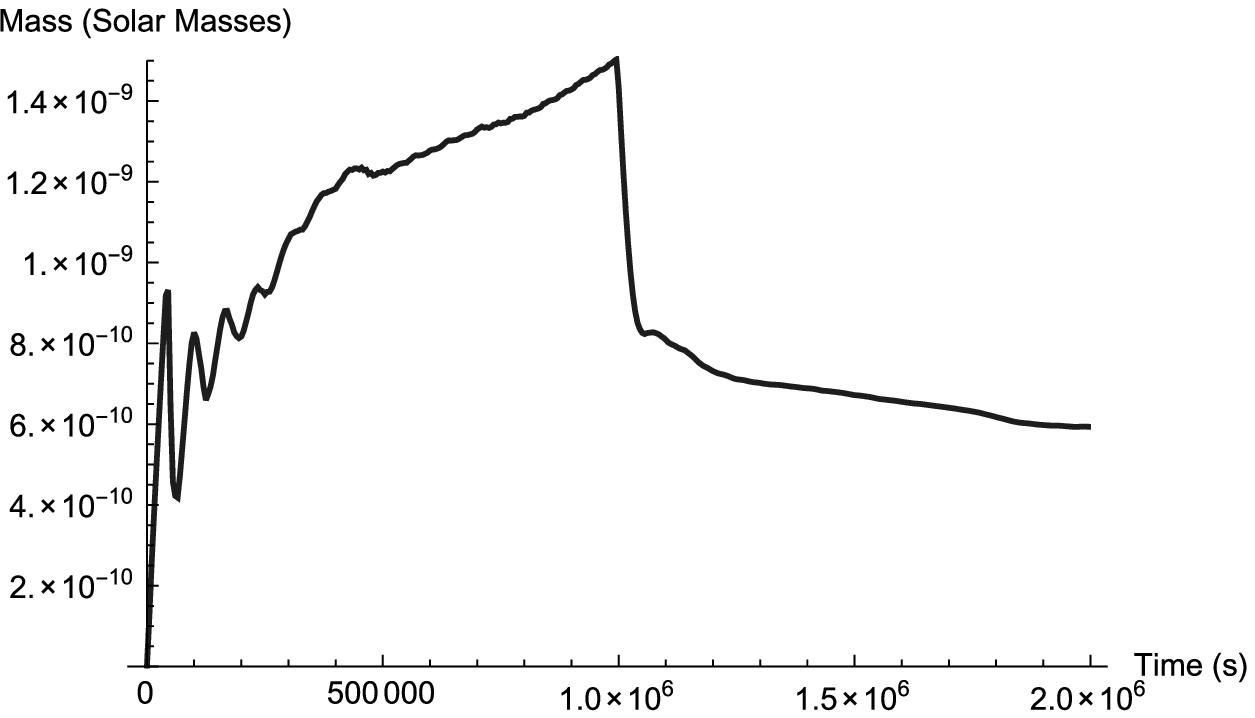}{rel}{Mass in the disk in $M_\odot$ as a function of time. Pulsations are turned off at 1 Ms and the disk is allowed to relax.}

\subsection{Variations on the +/- model}

As preliminary investigations into some possible future veins of research, we consider two variations on the +/- model. The first of these is to vary the ratio of the perturbation velocity to $\Delta v \equiv v_{orb}-v_{rot}$. Figure \ref{vodv} shows that decreasing this ratio proportionally decreases disk mass, while increasing it creates a much steeper increase, nearly a factor of ten, in disk mass. Understanding this strong sensitivity requires further study.

\articlefigure{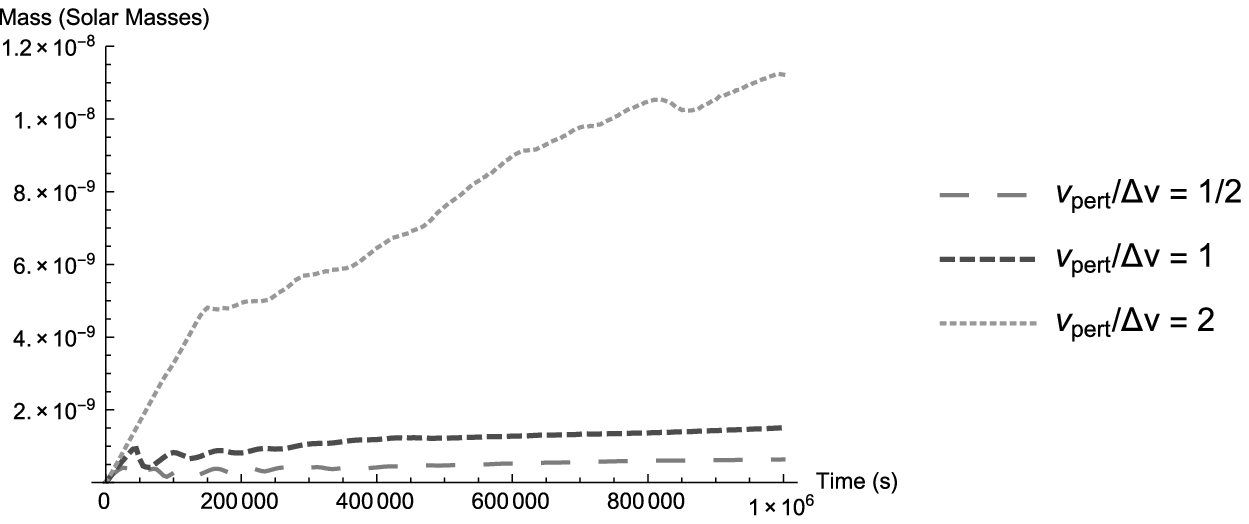}{vodv}{Mass in the disk in $M_\odot$ as a function of time for $v_\phi/\Delta v$ of 0.5, 1, and 2.}

The second variation on the base model, motivated by the complex multiperiodic nature of pulsations on $\mu$ Cen \citep{RivBaa98}, is to consider the effects of two perturbations beating against one another. For this, two perturbations with $v_{pert}=\Delta v/2$ are imposed with 10$\%$ separated periods ($P=40\;\mathrm{ks}\pm2\;\mathrm{ks}$), leading to a $400\;\mathrm{ks}$ beat period. Figure \ref{beat} shows that the disk mass oscillates with this beat period. During constructive interference, the mass peaks at a level comparable to that seen for a single mode mode; during destructive interference much of the material falls back on the star, much as occurred when the perturbations were turned off.

\articlefigure{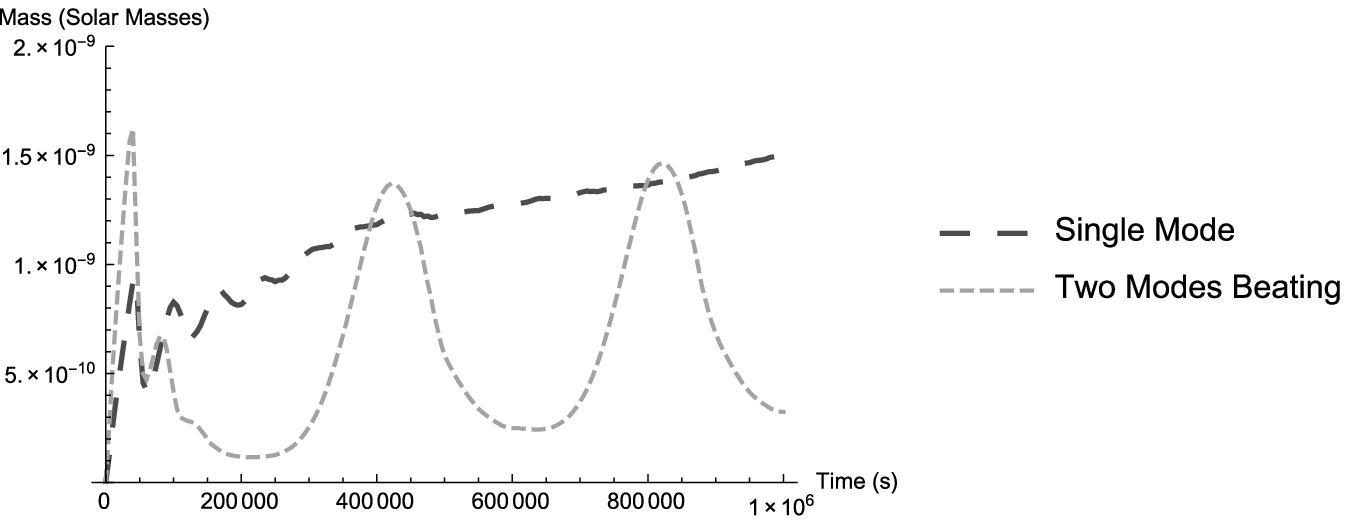}{beat}{Mass in the disk in $M_\odot$ as a function of time for a single mode pulsation and two pulsation modes beating against one another.}

\section{Conclusions and Future Work}\label{conclusions}

While preliminary, the results here provide a proof of concept for ejection of mass into the circumstellar environment by non-radial pulsations. They also demonstrate how such a disk, once generated, can persist in the absence of pulsations, viscous forces, and radiative forces. However, there is much room for future work. For instance, models should assume more moderate rotation rates, $v_{rot}=0.8-0.85v_{orb}$, which require $9-16$ times more energy to reach orbit. In addition, models should be extended to higher latitudes away from the equatorial plane using either 2D axisymmetry\footnote{Note that such an azimuthally symmetric mode would imply radial pulsation modes which are not observed in Classical Be stars but may still be considered meaningful as a computationally inexpensive first step.} or a full 3D model. Given the success of building and dissipating disks by viscous diffusion \citep{BjoCar05,CarBjo12}, viscous forces should be included in modeling the disk evolution. Given the high luminosity of the central star, one should also consider the effect of line driven forces in inhibiting the build-up of a disk, or ablating its surface layers. Nevertheless, the PDOME models here provide a promising step toward understanding the dynamics of Be disk formation, and a framework for such further investigations.


\acknowledgements We acknowledge funding support received from NASA ATP Grant NNX11AC40G, and helpful feedback from T. Rivinius during initial phases of this investigation.

\bibliography{kee_london}  

\appendix{Questions}

\emph{A. Okazaki: You mentioned that no viscosity is included in your simulations. But, your disks extend up to $~1.5\;R_*$, where the specific angular momentum is about 20$\%$ higher than at the stellar equatorial surface. What's the mechanism for angular momentum transport in your simulations?}

\vspace{10 pt}

\emph{N. D. Kee: It is correct that viscous forces are not explicitly included in these simulations. However, due to the use of a grid based code, the code is subject to numerical viscosity. This is likely sufficient to provide the radial extent which is seen in the results I have shown. However, confirming that this is the mechanism responsible for the spreading of the disk is still one of the issues still to be addressed.}

\end{document}